\documentclass{article}
\usepackage{dcase2025_techrep,amsmath,graphicx,url,times,booktabs, tabularx}

\title{ADAPTIVE KNOWLEDGE DISTILLATION USING A DEVICE-AWARE TEACHER FOR LOW-COMPLEXITY ACOUSTIC SCENE CLASSIFICATION}

\name{Seunggyu Jeong, Seongeun Kim\sthanks{corresponding author}}
\address{Seoul National University of Science and Technology, Seoul, South Korea, \\{wa975}@naver.com, sekim@seoultech.ac.kr
 }

\begin{document}

\ninept
\maketitle

\begin{sloppy}

\begin{abstract}
In this technical report, we describe our submission for Task 1, Low-Complexity Device-Robust Acoustic Scene Classification, of the DCASE 2025 Challenge. Our work tackles the dual challenges of strict complexity constraints and robust generalization to both seen and unseen devices, while also leveraging the new rule allowing the use of device labels at test time. Our proposed system is based on a knowledge distillation framework where an efficient CP-MobileNet student learns from a compact, specialized two-teacher ensemble. This ensemble combines a baseline PaSST teacher, trained with standard cross-entropy, and a 'generalization expert' teacher. This expert is trained using our novel Device-Aware Feature Alignment (DAFA) loss, adapted from prior work, which explicitly structures the feature space for device robustness. To capitalize on the availability of test-time device labels, the distilled student model then undergoes a final device-specific fine-tuning stage. Our proposed system achieves a final accuracy of 57.93\% on the development set, demonstrating a significant improvement over the official baseline, particularly on unseen devices.
\end{abstract}


\section{Introduction}
\label{sec:intro}

Acoustic Scene Classification (ASC) systems aim to categorize audio recordings into predefined scene classes. This field has gained prominence through the Detection and Classification of Acoustic Scenes and Events (DCASE) Challenge, which takes place annually \cite{1}. In this report, we describe our submission for Task 1 of the DCASE 2025 edition. This task concerns low-complexity, device-robust urban acoustic scene classification using the TAU Urban Acoustic Scenes 2025 Mobile dataset \cite{2}. It demands systems that not only perform well under stringent complexity constraints—a maximum of 128 kB of parameters and 30 million multiply-accumulate operations (MMACs)—but also exhibit strong generalization across diverse device conditions.

A key aspect of this year's challenge, differing from previous editions, is the availability of device labels during the testing phase. This allows for the development of systems that can adapt their inference process based on the known device type. Our methodology is designed to explicitly leverage this new rule, while also ensuring robustness for completely unseen devices.

Our approach is built upon the well-established Knowledge Distillation (KD) framework, which has consistently proven effective for training high-performance, low-complexity models \cite{3}. We employ a CP-MobileNet \cite{4} as our student model and utilize powerful Patchout FaSt Spectrogram Transformer (PaSST) \cite{5} models as teachers. Standard data augmentation techniques such as Mixup \cite{6} and Freq-MixStyle \cite{7,8} are also used to enhance overall robustness.

The main contribution of this work is a novel training framework that combines specialized knowledge distillation with an adaptive fine-tuning strategy. We distill knowledge from a compact teacher ensemble composed of two PaSST models: 1) a baseline teacher trained with standard cross-entropy loss, and 2) a 'generalization expert' teacher trained with our Device-Aware Feature Alignment (DAFA) loss. This loss, adapted from our prior work \cite{9}, explicitly structures the feature space for enhanced generalization. To capitalize on the new task rule, the distilled student model then undergoes a final device-specific fine-tuning stage. This step specializes the model's performance for the six known device types present in the training data, allowing for an adaptive inference strategy at test time. This combined approach is designed to achieve a robust performance-complexity trade-off across all device categories.

\section{SYSTEM COMPONENTS}
\label{sec:format}

\subsection{Student Network: CP-MobileNet}

For the student model, we selected CP-MobileNet (CPM) \cite{4}, an architecture that has demonstrated a remarkable balance between performance and computational cost in previous ASC tasks. It incorporates efficiency-focused designs, such as depth-wise separable convolutions inspired by MobileNets \cite{10}, to minimize its parameter count and computational load. Crucially, it retains architectural properties proven to be effective for ASC, most notably a carefully constrained receptive field which helps in learning discriminative local features from spectrograms.

To meet the challenge's complexity budget, our student model is a CP-MobileNet configured with 32 base channels, an expansion rate of 3, and a channel multiplier of 2.3. This configuration results in a model with approximately 128K parameters and 29.5 Million MACs, staying within the specified limits.

\subsection{Teacher Network: PaSST}

The foundation of our teacher models is the Patchout faSt Spectrogram Transformer (PaSST) \cite{5}, a powerful Transformer-based model. It employs a self-attention mechanism to effectively capture long-range temporal and spectral dependencies in audio spectrograms. Leveraging its pre-training on the large-scale AudioSet dataset \cite{11}, fine-tuned PaSST models have consistently set state-of-the-art benchmarks in various audio tasks, including ASC, and have proven to be exceptional knowledge sources for smaller CNN-based students.

\subsection{Data Augmentation Techniques}

To address the critical challenge of device mismatch, we employ a combination of three data augmentation techniques during training:
\begin{itemize}
    \item \textbf{Freq-MixStyle}\cite{7,8}: This technique combats device mismatch by swapping the statistics of frequency-band-level features between different training samples, encouraging the model to learn representations that are robust to channel variations.
    \item \textbf{Mixup}\cite{6}: We utilize Mixup, which generates new training samples by creating convex combinations of pairs of samples and their corresponding labels. Mixup serves as a strong form of regularization, improving model generalization.
\end{itemize}

\section{TRAINING METHODOLOGY}
\label{sec:pagestyle}

Building upon the components described in the previous section, our core contribution lies in a novel training methodology. This section details our training process, which is centered on a specialized ensemble knowledge distillation (KD) framework designed to enhance model generalization. We will describe the composition of our unique teacher ensemble and the overall loss function used to train the student model.

\subsection{Specialized Ensemble Distillation Framework}
Our training paradigm is centered around a knowledge distillation (KD) framework that utilizes a compact, yet powerful, specialized teacher ensemble. Instead of separate strategies for different device conditions, we employ a single, unified distillation process to train our student model.

The teacher ensemble consists of two PaSST models, which provide a combined supervisory signal for the student:
\begin{itemize}
    \item \textbf{A Baseline Teacher}: A PaSST model trained conventionally using only the standard cross-entropy loss. This teacher provides a strong and accurate signal based on the primary task of scene classification.
    \item \textbf{A Generalization Expert Teacher}: A second PaSST model trained with a composite loss function. This loss combines the standard cross-entropy loss with our Device-Aware Feature Alignment (DAFA) loss, which is detailed in Section \ref{sec:dfa_loss}. This teacher specializes in creating a structured and device-robust feature representation.
\end{itemize}
The soft predictions from these two teachers are averaged to form the final distillation target. After the main distillation process is complete, the student model undergoes a final device-specific fine-tuning step to further optimize its performance on the characteristics of known devices.

\subsection{Knowledge Distillation Loss}
The student model is trained by minimizing a composite loss function that combines a standard cross-entropy loss for hard labels and a distillation loss for soft teacher predictions. The total loss $\mathcal{L}$ is defined as:
\begin{equation}
    \mathcal{L} = (1-\lambda)\mathcal{L}_{CE}(\delta(\mathbf{z}_S), \mathbf{y}) + \lambda \tau^2 \mathcal{L}_{KL}(\delta(\mathbf{z}_S/\tau), \delta(\mathbf{z}_T/\tau))
\end{equation}
where $\mathbf{z}_S$ and $\mathbf{z}_T$ are the student and teacher logits. $\mathbf{y}$ is the one-hot encoded ground truth label. The hyperparameter $\lambda$ balances the two loss terms, and the temperature $\tau$ softens the probability distributions from the softmax activation $\delta$. The teacher logits $\mathbf{z}_T$ in this equation represent the averaged logits from the two-teacher ensemble described in the previous subsection.

\begin{table}[t]
\centering
\caption{Teacher accuracy (\%) based on different configurations.}
\label{tab:ensemble_ablation}
\begin{tabular}{l|cccc}
\hline
\textbf{Teacher Config} & \textbf{Overall} & \textbf{Real} & \textbf{Seen} & \textbf{Unseen} \\ \hline
\textit{Single Teacher} & & & & \\
\quad $T_{CE1}$ & 59.78 & 65.22 & 56.77 & 57.36 \\
\quad $T_{CE2}$ & 59.22 & 64.97 & 56.10 & 56.60 \\
\quad $T_{DAFA}$ & 59.00 & 64.70 & 55.60 & 56.71 \\ \hline
\textit{Two-Teacher Ensemble} & & & & \\
\quad $T_{CE1} + T_{CE2}$ & \textbf{60.73} & \textbf{66.47} & 57.41 & 58.31 \\
\quad $\mathbf{T_{CE1} + T_{DAFA}}$ & \textbf{60.73} & 66.12 & \textbf{57.73} & \textbf{58.35} \\ \hline
\end{tabular}
\end{table}

\begin{table*}[t]
\centering
\caption{Final accuracy (\%) by device on the validation set.}
\label{tab:final_results_horizontal}
\resizebox{\textwidth}{!}{%
\begin{tabular}{l|ccccccccc|c}
\hline
\textbf{Stage} & \textbf{A} & \textbf{B} & \textbf{C} & \textbf{S1} & \textbf{S2} & \textbf{S3} & \textbf{S4} & \textbf{S5} & \textbf{S6} & \textbf{Overall} \\ \hline
After Distillation & 64.33 & 52.43 & 57.33 & 51.39 & 51.70 & 54.58 & 54.67 & 55.12 & 50.00 & 54.61 \\
After DSFT (Final) & 69.82 & 60.67 & 64.80 & 53.55 & 54.94 & 57.82 & 54.67 & 55.12 & 50.00 & 57.93 \\ \hline
\textbf{Improvement ($\Delta$)} & \textbf{+5.49} & \textbf{+8.24} & \textbf{+7.47} & \textbf{+2.16} & \textbf{+3.24} & \textbf{+3.24} & \textbf{-} & \textbf{-} & \textbf{-} & \textbf{+3.32} \\ \hline
\end{tabular}%
}
\end{table*}

\section{DEVICE-AWARE FEATURE ALIGNMENT LOSS}
\label{sec:dfa_loss}

Our Device-Aware Feature Alignment (DFA) loss is introduced as a regularization term, added to the primary cross-entropy objective ($\mathcal{L}_{CE}$), to train our generalization expert teacher. The purpose of this loss is to structure the embedding space to be robust against device variations. The DFA loss itself consists of two weighted components: a Device Cohesion-Separation Loss (DCSL) that manages device-specific clusters, and a Global Device Alignment Loss (GDAL) that provides overall structural coherence. The total objective for the expert teacher is:

\begin{equation}
    \mathcal{L}_{total} = \mathcal{L}_{CE} + \lambda_{dcsl}\mathcal{L}_{DCSL} + \lambda_{gdal}\mathcal{L}_{GDAL}
\end{equation}
Through empirical validation, we set the weights to $\lambda_{dcsl} = 0.01$ and $\lambda_{gdal} = 0.01$.

\subsection{Device Cohesion-Separation Loss (DCSL)}
The DCSL component is designed to improve the separability of device-specific feature clusters. The numerator of this ratio, the intra-device scatter ($S_W$), quantifies the compactness of features within each device cluster. The denominator, the inter-device scatter ($S_B$), measures the separation between the centroids of different device clusters. The loss is formulated as:

\begin{equation}
    \mathcal{L}_{DCSL} = \frac{S_W}{S_B + \epsilon}
\end{equation}
Minimizing this objective simultaneously encourages each device cluster to become internally tight while pushing the clusters externally apart from each other, thus preserving device-specific information in a structured manner.

\subsection{Global Device Alignment Loss (GDAL)}
The GDAL component acts as a global regularizer on the feature space, preventing the device-specific clusters shaped by DCSL from fragmenting the embedding space. It penalizes the deviation of individual device centroids ($\mu_d$) from a global centroid ($\mu_G$), which represents the mean of all features in a batch. By tethering all device clusters to this common anchor point, GDAL ensures that the model learns a coherent, shared representation space. This is crucial for the model's ability to generalize to unseen devices. The loss is defined as the mean squared distance from the device centroids to the global centroid:

\begin{equation}
    \mathcal{L}_{GDAL} = \frac{1}{|D|} \sum_{d \in D} ||\mu_d - \mu_G||_2^2
\end{equation}

\section{EXPERIMENTS AND RESULTS}
\label{sec:experiments}

In this section, we present the experimental validation of our proposed system. We first describe the experimental setup, then analyze the performance based on different teacher ensemble compositions to identify the optimal configuration. Finally, we present the detailed performance of our final proposed system, highlighting the effect of device-specific fine-tuning.

\subsection{Experimental Setup}
Our experiments are conducted on the TAU Urban Acoustic Scenes 2025 Mobile development dataset \cite{2}, which contains 1-second audio clips from 10 acoustic scenes, recorded with a variety of real devices. We follow the official training and validation splits provided by the challenge organizers. The primary evaluation metric is the log loss, and we also report accuracy (\%) for interpretability.
The main knowledge distillation stage runs for 500 epochs using the Adam optimizer with a batch size of 128 and a peak learning rate of 5e-4, managed by a scheduler with linear warmup and cosine decay. For distillation, we use a temperature $\tau=2$ and a loss weight $\lambda=0.98$. After distillation, the device-specific fine-tuning stage is performed for an additional 100 epochs for each of the 6 known devices, using a lower learning rate of 1e-5.

\subsection{Analysis of Teacher Ensemble Composition}
To determine the optimal teacher ensemble for our knowledge distillation framework, we conducted a series of experiments. We first distilled knowledge into the student model from three individual teachers: two baseline teachers trained only with Cross-Entropy loss on different seeds ($T_{CE1}$, $T_{CE2}$), and one expert teacher trained with both Cross-Entropy and our DAFA loss ($T_{DAFA}$). Subsequently, we evaluated various two-teacher ensemble configurations.

The results in Table \ref{tab:ensemble_ablation} provide several interesting insights. Among the single teachers, the baseline teacher $T_{CE1}$ achieved a slightly higher overall performance than the $T_{DAFA}$ expert. This suggests that for a single knowledge source, a standardly trained model provided a strong foundation for the student.

However, the true benefit of our approach becomes evident in the ensemble results. Both two-teacher ensembles significantly outperform any single teacher, confirming the general effectiveness of ensembling. Comparing the two ensembles, we observe that they achieve the exact same overall accuracy of 60.73\%. Despite this tie, the $\mathbf{T_{CE1} + T_{DAFA}}$ ensemble demonstrates a clear advantage in generalization. It achieves the highest accuracy on both the Seen (57.73\%) and, more critically, the Unseen (58.35\%) device subsets. This indicates that including the DAFA expert in the ensemble successfully regularizes the student model and boosts its robustness to novel device characteristics, which is a primary goal of this work. Therefore, based on its superior generalization performance, we selected the ($T_{CE1} + T_{DAFA}$) ensemble for our final proposed system.

\subsection{Final System Performance}
This section presents the detailed results of our final proposed system, which employs the 2 teacher ensemble for distillation, followed by a device-specific fine-tuning (DSFT) stage. Table \ref{tab:final_results_horizontal} shows the per-device accuracy before and after the DSFT stage to clearly demonstrate its impact.

The results highlight two key aspects of our system. First, the model already achieves strong performance after the main distillation phase, particularly on unseen devices, thanks to the powerful teacher ensemble. Second, the DSFT stage provides a substantial and consistent performance boost across all six known devices present in the training data, with an average improvement of 3.32\%. This confirms that our fine-tuning strategy is highly effective at specializing the model, successfully leveraging the availability of device labels at test time as per the challenge rules.


\section{CONCLUSION}
\label{sec:conclusion}

In this technical report, we presented our system for the DCASE 2025 Task 1, addressing the challenge of low-complexity, device-robust acoustic scene classification. Our proposed method is centered on a knowledge distillation framework that leverages a compact, specialized two-teacher ensemble to train an efficient CP-MobileNet student. The core of our approach lies in this ensemble, which combines a baseline teacher with a 'generalization expert' trained using our novel Device-Aware Feature Alignment (DAFA) loss to instill robust, device-agnostic features. Furthermore, we effectively capitalized on the new task rule by applying a final device-specific fine-tuning stage, which adapts the model to known device characteristics using test-time device labels.

Experimental results on the development set validate our approach, achieving a final accuracy of 57.93\%. Our analyses confirmed that the specialized teacher ensemble was crucial for improving generalization to unseen devices, while the adaptive fine-tuning stage consistently and significantly boosted performance on all known devices. This demonstrates that our two-stage strategy—first instilling general robustness via specialized distillation, and then adapting to specifics via targeted fine-tuning—is a powerful and effective solution for the complex device generalization problem presented in this year's challenge.

\bibliographystyle{IEEEtran}
\bibliography{refs}
%
%
%
%
%
%
%
%
%

\end{sloppy}
\end{document}